\documentclass[aps,prd,twocolumn,10pt,superscriptaddress,amsmath,amssymb,longbibliography]{revtex4-1}

\usepackage{graphicx}
\usepackage{subfigure}
\usepackage{epstopdf}
\usepackage{hyperref}
\usepackage{color}
\usepackage{bbold}

\newcommand{\unitvec}{\hat{\mathbf{e}}}

\newcommand{\sigmab}{\boldsymbol{\sigma}}
\newcommand{\magn}{\mathbf{m}}
\newcommand{\pos}{\mathbf{r}}

\begin{document}

\title{Majorana bound states in magnetic skyrmions imposed onto a superconductor}

\author{Stefan Rex}
\affiliation{Institut f\"ur Nanotechnologie, Karlsruhe Institute of Technology, 76021 Karlsruhe, Germany}
\affiliation{\mbox{Institut f\"ur Theorie der Kondensierten Materie, Karlsruhe Institute of Technology, 76128 Karlsruhe, Germany}}
\author{Igor V. Gornyi}
\affiliation{Institut f\"ur Nanotechnologie, Karlsruhe Institute of Technology, 76021 Karlsruhe, Germany}
\affiliation{\mbox{Institut f\"ur Theorie der Kondensierten Materie, Karlsruhe Institute of Technology, 76128 Karlsruhe, Germany}}
\affiliation{A. F. Ioffe Physico-Technical Institute, 194021 St. Petersburg, Russia}
\author{Alexander D. Mirlin}
\affiliation{Institut f\"ur Nanotechnologie, Karlsruhe Institute of Technology, 76021 Karlsruhe, Germany}
\affiliation{\mbox{Institut f\"ur Theorie der Kondensierten Materie, Karlsruhe Institute of Technology, 76128 Karlsruhe, Germany}}
\affiliation{L. D. Landau Institute for Theoretical Physics RAS, 119334 Moscow, Russia}
\affiliation{Petersburg Nuclear Physics Institute, 188300 St. Petersburg, Russia}

\begin{abstract}
We consider a superconducting film exchange-coupled to a close-by chiral magnetic layer and study how magnetic skyrmions can induce the formation of Majorana bound states (MBS) in the superconductor. Inspired by a proposal by Yang \textsl{et al.} [Phys. Rev. B 93, 224505 (2016)], which suggested MBS in skyrmions of even winding number, we explore whether such skyrmions could result from a merger of ordinary skyrmions. We conclude that the formation of higher-winding skyrmions is not realistic in chiral magnets. Subsequently, we present a possibility to obtain MBS from realistic skyrmions of winding number one, if a skyrmion-vortex pair is formed instead of a bare skyrmion. Specifically, we show that MBS are supported in a pair of a circular skyrmion and a vortex which both have a winding number of one. We back up our analytical prediction with results from numerical diagonalization and obtain the spatial profile of the MBS. In light of recent experimental progress on the manipulation of skyrmions, such systems are promising candidates to achieve direct spatial control of MBS.
\end{abstract}

\maketitle

\section{Introduction}\label{Sec:Intro}
The search for Majorana bound states (MBS) in superconducting heterostructures has experienced massive theoretical and experimental efforts in recent years. The enormous interest in such MBS is owed to their intriguing properties, most notably the topological protection against local perturbations and the non-Abelian exchange statistics \cite{Ali12, Ben13}, which have fueled ambitions to use MBS as constituents of topological qubits in quantum computers \cite{StL13, NSS08}.

By now, the most prominent platform for Majorana physics are semiconductor-superconductor hetero-nanowires \cite{Kit01, LSD10, ORO10}, where a combination of strong spin-orbit coupling (SOC), superconductivity, and a Zeeman field yields an effective $p$-wave order parameter and a topological phase. Starting from the first measurements in 2012 \cite{MZF12, RLF12, DRM12, DYH12}, increasingly appealing experimental indications for the existence of MBS could be gained with improved nanowire architectures, as, e.g., in Refs.~\onlinecite{DVH16, GZB18}. What remains difficult, though conceptionally possible, is to realize real-space braiding in nanowire-based systems.

Branching off from such nanowires, also linear arrangements of magnetic adatoms on superconductors (Shiba chains) have been studied \cite{CEA11, NDB13, PGO13, KSY13, BrS13, VaF13, BSH15, PPG15, CSF16, SFC16}, in which the internal magnetization replaces the external field. Furthermore, if the magnetic order in a chain or wire is helical rather than ferromagnetic, the winding of the magnetization induces an effective SOC, thus also replacing the need for intrinsic high-SOC materials. This interesting approach was soon expanded theoretically to two-dimensional Shiba lattices \cite{NTN13, PWR14, RoO15, LNW16}, where non-collinear magnetic phases were shown to lead to a rich topological phase diagram. Signatures of a chiral topological phase have been measured recently in ferromagnetic islands on a superconductor \cite{MMB18}.

In principle, similar ideas can be applied to bilayer systems where a superconducting substrate is covered by a thin magnetic film rather than decorated with isolated magnetic impurities. Non-collinear phases in magnetic films constitute by themselves a very active and fruitful field of research \cite{HFT17}. In particular, chiral magnets with Dzyaloshinskii-Moriya interaction \cite{Dzy58, Mor60} can exhibit phases with stable magnetic skyrmions \cite{MBJ09, YOK10, HBM11}. Skyrmions are particle-like textures characterized by a topological winding number $n$ \cite{NaT13}, which is defined by
\begin{equation}\label{Eq:WindingNumber}
n = \int\frac{\textrm{d}^2r}{4\pi}\magn\cdot\left(\partial_x\magn\times\partial_y\magn\right)
\end{equation}
(in terms of the unit vector $\magn$). They give rise to emergent electrodynamics \cite{SRB12} and have been studied intensely, in particular, in context of potential applications in spintronics devices \cite{Wie16}.

Given the technological maturity in the creation and control of skyrmions \cite{FBT16, FRC17, EMR18}, it seems natural to investigate in how far they can, if combined with superconductivity \cite{VSM18}, serve as carriers for MBS. However, relatively little work on magnetic skyrmions on superconductors has been done so far. For skyrmions on $s$-wave and $p$-wave superconductors, Yu-Shiba-Rusinov-like bound states have been considered \cite{PNB16, PWP16}. With regard to Majorana physics, effective $p$-wave phases with chiral edge modes have been predicted \cite{ChS15, MCR18}, whereas localized MBS may emerge in elongated skyrmions \cite{GSK18} which resemble nanowires. For more complicated skyrmion textures, Yang \textsl{et al.} \cite{YSK16} have derived the criterion that a pair of MBS can form for skyrmions of even winding number, with one state being localized at the skyrmion core and the partner state at the outer rim of the skyrmion. We will return to this criterion in more detail later.

The idea of having an MBS captured inside a skyrmion is very appealing, because the existing tools for skyrmion manipulation could potentially be employed to navigate single Majorana modes. It is particularly noteworthy that a recent experiment \cite{HDL18} has demonstrated the possibility to move individual skyrmions in the plane with a scanning tunneling microscope tip. In that way, one could envision explicit real-space braiding operations with MBS.

The goal of this paper is to point out a feasible way to generate MBS from magnetic skyrmions. Importantly, the required skyrmions must have a realistic spatial shape. We take the proposal \cite{YSK16} of MBS in skyrmions of even winding number, in particular double-winding skyrmions (DWS), as a starting point. Skyrmions in chiral magnets are, in contrast, characterized by a winding number $n=1$ (perhaps $n=-1$), which we refer to as single-winding skyrmions (SWS). In the first part of our work, Sec.~\ref{Sec:Merger}, we investigate whether DWS could be formed by the fusion of two SWS in a chiral magnetic film, as the conservation of the total winding number in the system would insinuate. We show that, when a skyrmion merger is enforced, the destruction of one skyrmion is energetically favorable compared to the protection of the topological invariant. We note, though, that DWS may occur in systems where skyrmion formation is driven by different mechanisms, as dipolar interactions or frustration effects \cite{YTK14, LeM15, ZZE16, OHM17, HaM19}, which we do not study in this paper.

In the second part, Secs.~\ref{Sec:Criterion} and \ref{Sec:Results}, we generalize the criterion for MBS presented in Ref.~\onlinecite{YSK16} in order to cirumvent the requirement of a DWS, by combining an SWS with a vortex in the superconductor. Such skyrmion-vortex pairs have been shown to be stable under suitable conditions \cite{HSR16}. MBS being bound to vortex cores is a concept that has been known for many years \cite{Iva01} for two-dimensional $p$-wave superconductors. Topological superconductivity in two dimensions emerges, for instance, at topological insulator-superconductor interfaces \cite{FuK08, LTY10}, and signatures of MBS at vortices have indeed been reported \cite{SZH16, SuJ17, LCZ18}. Bare vortices are, however, not easily individually controllable. In this regard, the skyrmion-vortex pairs that we address allow one to envision more flexible applications in quantum technology.

\section{Merger of single-winding skyrmions in chiral magnets}\label{Sec:Merger}
In this section, we investigate a merger of two SWS as a potential way to generate a DWS in a chiral magnet, having in mind the proposal in Ref.~\onlinecite{YSK16}. For simplicity, we do not include the proximitized superconductor at this point, because the relevant energy scale in the magnet, given by the exchange interaction, will be large compared to the induced superconductivity in the magnet.

The texture of a circular skyrmion can be expressed as
\begin{equation}\label{Eq:SkyrmionTexture}
\magn_\text{sk}(r, \varphi) = \begin{pmatrix}
                      \cos f(\varphi)\sin g(r) \\						\sin f(\varphi)\sin g(r) \\						\cos g(r)
                    \end{pmatrix}
\end{equation}
in polar coordinates $(r,\varphi)$ relative to the skyrmion center, where $f(\varphi)$ and $g(r)$ denote the angular and radial profiles, respectively. In general, the winding number~\eqref{Eq:WindingNumber} is
\begin{equation}
n = \left[\frac{f(\varphi)}{2\pi}\right]_0^{2\pi}\left[\frac{-\cos g(r)}{2}\right]_0^\infty
\end{equation}
and one has to impose the boundary conditions $g(0)=\pi$ and $g(\infty)=0$. We assume an exponential radial profile, $g(r)=\pi e^{-r/R}$ with the skyrmion radius $R$, and the angular profile $f(\varphi)=n\varphi + \varphi_0$. If $n=1$, the offset angle $\varphi_0$ indicates N\'eel type ($\varphi_0=0,\pi$), Bloch type ($\varphi=\pm\pi/2$), or any intermediate configuration (all of which are topologically equivalent).

The energy of the magnetic configuration on a lattice is given by
\begin{eqnarray}\label{Eq:MagEnergy}
H &=& -J\sum_{\left<ij\right>}\magn_i\cdot\magn_j - \sum_{\left<ij\right>} \mathbf{D}_{ij}\cdot\left(\magn_i\times\magn_j\right) \notag\\
&&{}+ K\sum_i\left(M_i\cdot\unitvec_z\right)^2 + \mu\sum_i\mathbf{B}\cdot\magn_i \,,
\end{eqnarray}
where the first term describes exchange coupling of nearest-neighbor sites, the second term is the Dzyaloshinskii-Moriya interaction \cite{Dzy58, Mor60}, the third term indicates easy-plane anisotropy if $K>0$, %
and the last term accounts for an external field. In Eq.~\eqref{Eq:MagEnergy}, $\magn_i$ denotes the unit vector parallel to the magnetic moment $\mathbf{M}_i=\mu\magn_i$ at site $i$.

\begin{figure}
\includegraphics[width=\columnwidth]{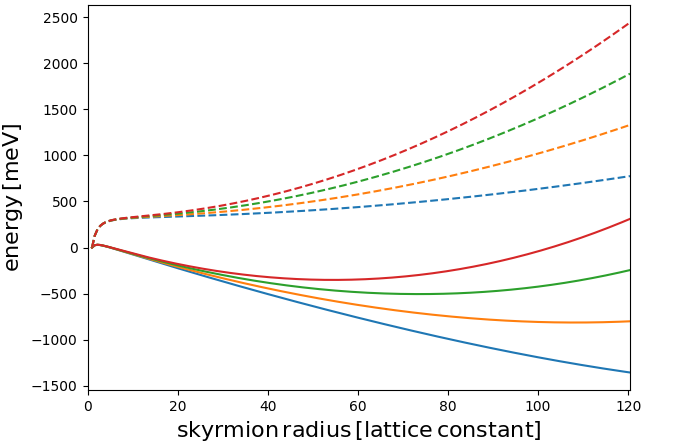}
\caption{\label{Fig:EnergyRadius}Energy of a SWS (solid) or DWS (dashed) as a function of its size under an external out-of-plane field of $100\,\mathrm{mT}$ (blue), $150\,\mathrm{mT}$ (orange), $200\,\mathrm{mT}$ (green), and $250\,\mathrm{mT}$ (red). Equation~\eqref{Eq:MagEnergy} was implemented on a triangular lattice with 1386$\times$1600 sites and $J=13.1\,\mathrm{meV}$, $|\mathbf{D}|=0.6\,\mathrm{meV}$, $K=0.005\,\mathrm{meV}$, and $\mu=1.8\mu_B=0.1042\,\mathrm{meV/T}$. For comparability, each graph was shifted by a constant energy offset proportional to $B$.}
\end{figure}

\begin{figure*}[t]
\includegraphics[width=0.9\textwidth]{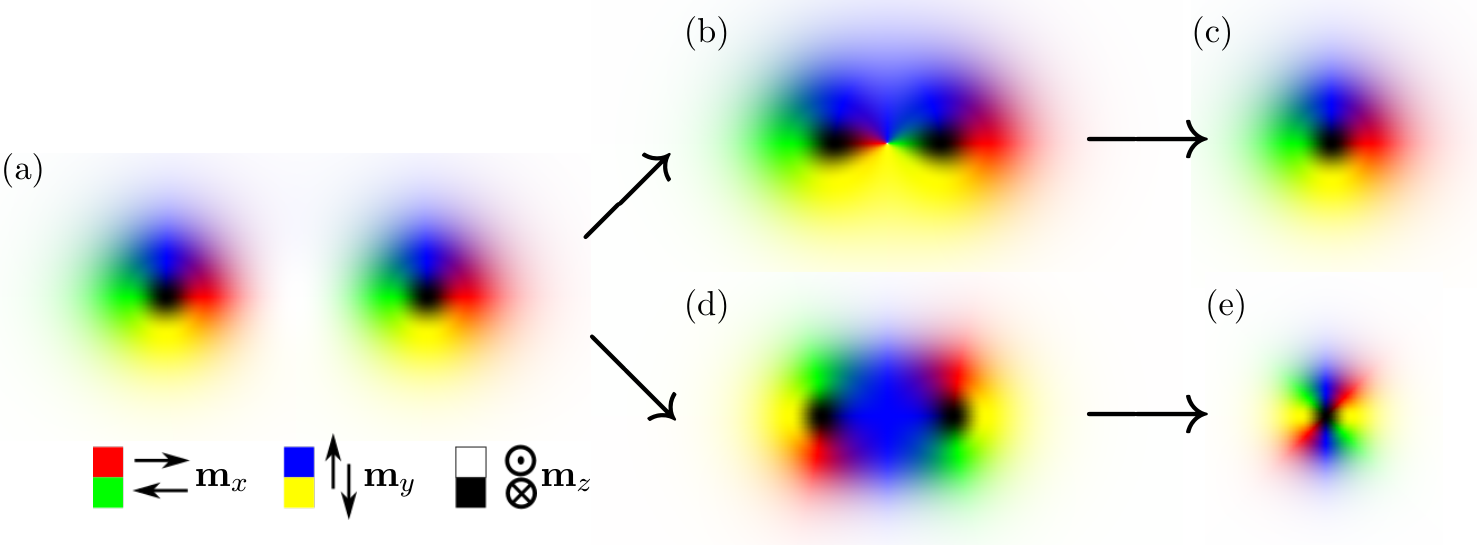}
\caption{\label{Fig:Merger}Two SWS (a) can be merged either undergoing a pointlike discontinuity as shown in (b), resulting in a single SWS (c), or in continuous fashion (d), resulting in one DWS (e). Colors indicate the in-plane component of the magnetization, brightness the out-of-plane component.}
\end{figure*}

The Dzyaloshinskii-Moriya interaction favors a relative canting of neighboring magnetic moments and thereby enriches the phase diagram as compared to non-chiral magnets. While the ground state at $B=0$ typically exhibits spiral magnetic order, a skyrmion phase may occur at intermediate field strengths followed by collinear alignment at higher fields. Here, we take $\mathbf{D}_{ij}$ to be in-plane and orthogonal to the $\left<ij\right>$ bond (favoring N\'eel type winding). %
We choose a set of parameters in Eq.~\eqref{Eq:MagEnergy} which resembles a Co/Ru(0001) bilayer system \cite{HDL18} where stable SWS exist.
This is illustrated in Fig.~\ref{Fig:EnergyRadius}, where the energy of SWS and DWS is shown as a function of the skyrmion radius for multiple field strengths. An SWS has an energy minimum at an optimal radius for suitable field strengths. The DWS, in contrast, never develops such a minimum.
This is understood as follows: Whenever $n\geq2$, the rotation of the magnetization along the radial direction continuously varies with $\varphi$ between the preferred N\'eel type (where $\mathbf{D}_{ij}\cdot(\magn_i\times\magn_j)>0$), the neutral Bloch type ($\mathbf{D}_{ij}\cdot(\magn_i\times\magn_j)=0$), and the unfavorable (opposite) N\'eel type ($\mathbf{D}_{ij}\cdot(\magn_i\times\magn_j)<0$). In total, these contributions cancel  if $f$ is linear in $\varphi$ [see Appendix~\ref{App:DMEnergy} for a calculation of the Dzyaloshinskii-Moriya energy of a skyrmion with the general shape given by Eq.~\eqref{Eq:SkyrmionTexture}]. Although DWS do not form spontaneously, their energy profile is relatively flat at not too large $B$, such that a DWS might still have a sufficient lifetime if it can be created artificially, or even persist as a metastable state if the external field is switched off after the initial nucleation of skyrmions. Therefore we now turn to the DWS creation by a skyrmion merger.

When two SWS are merged, there are two conceivable scenarios: (i) a discontinuous process in which the overall winding number is allowed to change and (ii) the continuous merger resulting in a DWS due to the topological protection of $n$. Both cases are depicted in Fig.~\ref{Fig:Merger}. In (i), a pointlike discontinuity appears, similar to the case studied in Ref. \onlinecite{MKS13}, at the critical skyrmion distance $d_0=2R\text{ln}3$ [with our specific choice of $g(r)$]. The discontinuity can also be interpreted as an emergent antiskyrmion~\cite{TaF14}. To avoid this discontinuity in (ii), the textures of both SWS must be rotated such that their contributions cannot cancel in the overlap region. 

\begin{figure}[t]
\includegraphics[width=\columnwidth]{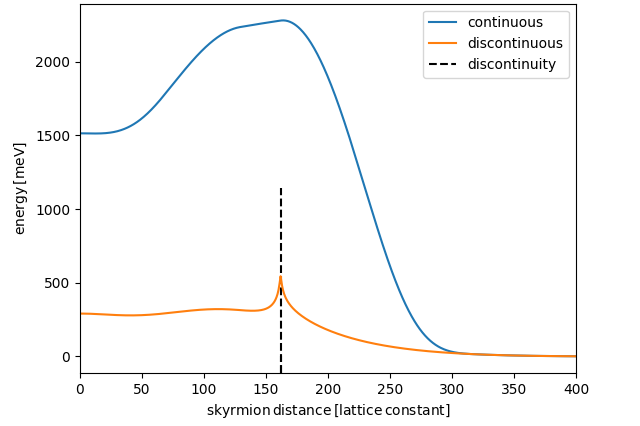}
\caption{\label{Fig:Merger-Energy}Energy evolution in a continuous (blue) and discontinuous (orange) SWS merger processes with the same parameters as in Fig.~\ref{Fig:EnergyRadius} at $B=200\,\text{mT}$ on a triangular lattice with $4158\times1600$ sites. The SWS radius $R=73.6a$ corresponds to the optimum in Fig.~\ref{Fig:EnergyRadius}.}
\end{figure}

It is clear that both processes involve an energy barrier. Therefore, any such process has to be enforced externally, e.g., by moving the SWS with a tunneling tip \cite{HDL18}. In order to decide  whether (i) or (ii) would be energetically favorable, we have implemented a specific instance of a continuous and a discontinuous merger (the ones shown in Fig.~\ref{Fig:Merger}), starting from the same initial configuration. Details on the magnetic texture during the merger can be found in Appendix~\ref{App:Merger}. By means of Eq.~\eqref{Eq:MagEnergy}, the energy can be tracked throughout each process. We present the result in Fig.~\ref{Fig:Merger-Energy}. Note that the sections displayed in Fig.~\ref{Fig:Merger} do not represent the full size of the system on which the calculations were performed. We use the same parameters as in Fig.~\ref{Fig:EnergyRadius} and $R=73{.}6a$, corresponding to the energy minimum at $B=200\,\text{mT}$.

The main observation is that the barrier related to the discontinuity in (i) is small compared to the energy required in (ii). The reason is, qualitatively, that the rotation of the SWS textures in (ii) acts against the Dzyaloshinskii-Moriya term in the entire area of each SWS. In contrast, the ``collision area'' in (i) in the vicinity of the discontinuity is comparably small, containing only a few sites at which the exchange energy becomes large. To be more precise, the energy related to the Dzyaloshinskii-Moriya interaction is a numerical constant times $D(R/a)\cos\varphi_0$ (see Appendix~\ref{App:DMEnergy}), where $\varphi_0$ changes from $0$ to $\pm\frac{\pi}{2}$ in going from panel (a) to panel (d) in Fig.~\ref{Fig:Merger}. For our parameters, this corresponds to $1600\,\text{meV}$ per skyrmion. This is consistent with the barrier height in Fig.~\ref{Fig:Merger-Energy}, given that the overlap of the skyrmions is not taken into account in our estimate. In the discontinuous merger, on the other hand, the barrier height is determined by the exchange energy in an area $A$ around the discontinuity at $d_0$ in which $\magn$ changes its direction. This area scales as $A\propto d_0^2$. The exchange energy density can be expressed as $\propto J(\nabla\magn)^2$, where the derivative of $\magn$ scales with $1/d_0$. Hence, the exchange energy contribution from the vicinity of the discontinuity is a constant proportional to $J$ which does not scale with the skyrmion size.

In this section, we did not account for the proximitized superconducting layer. In principle, the superconductor can lead to additional energy contributions in the merger process. For instance, bound states similar to Yu-Shiba-Rusinov states exist in the SWS \cite{PNB16, PWP16} and can induce antiferromagnetic interactions \cite{YGD14}. The energy scale of such effects is proportional to the superconducting gap. While these interactions can have important consequences on magnetic impurities on a superconductor (where they compete only with the Ruderman-Kittel-Kasuya-Yosida interaction), the \emph{direct} magnetic exchange interaction of strength $J$ will be dominant in the case of a magnetic film covering the superconductor. Referring again to a Co/Ru(0001) bilayer, the superconducting transition temperature of ruthenium is $T_c=470\,\text{mK}$ \cite{HuG57}, corresponding to $\Delta(T=0)\approx \frac{\pi}{e^\gamma}k_BT_c = 0.07\,\text{meV}$, which is negligible compared to $J=13.1\,\text{meV}$ \cite{HDL18}.

In total, we find that in an enforced SWS merger, the chiral magnetic film would favor the deletion of a skyrmion -- notably despite its topological nature -- over the formation of a DWS. Although our specific implementations of the two scenarios do not represent fully optimized paths, the result in Fig.~\ref{Fig:Merger-Energy} demonstrates the generic energy proportions, because the large discrepancy between small-area versus large-area energy cost cannot be overcome by small adjustments. The only possible exception would be extremely small skyrmions, which seem, however, unfavorable to localize a MBS. In essence, this rules out the possibility to generate MBS by DWS creation in the known Dzyaloshinskii-Moriya-driven skyrmion host materials. We turn therefore to cases where MBS are obtained from SWS in the following sections.

\section{Majorana criterion in a skyrmion-vortex pair}\label{Sec:Criterion}
Here, we present a generalization of the argument given in~\onlinecite{YSK16} to circumvent the necessity of DWS. Namely, we consider not a bare skyrmion on a superconductor, but the composite object of a skyrmion and a vortex in the superconductor \cite{HSR16}, thus incorporating the phase winding of the superconducting order parameter in addition to the skyrmion winding number. A similar argument solely based on the effect of phase winding was recently presented for superconducting hollow cylinders~\cite{LWH18, VDK18}. Consider the Hamiltonian $H=\int d^2r \Psi^\dagger(r)\mathcal{H}\Psi(r)$ with
\begin{equation}\label{Eq:Hamiltonian}
\mathcal{H} = -\!\left(\frac{1}{2m}\nabla^2+\mu\right)\!\tau_z + \lambda\magn_\text{sk}(\pos)\cdot\sigmab + \text{Re}(\Delta)\tau_x - \text{Im}(\Delta)\tau_y
\end{equation}
in Nambu space, $\Psi=(c_\uparrow, c_\downarrow, c^\dagger_\downarrow, -c^\dagger_\uparrow)$, for electrons with effective mass $m$ at a chemical potential $\mu$ exchange-coupled to the magnetic skyrmion texture $\magn_\text{sk}(\pos)$, in the presence of a superconducting $s$-wave gap $\Delta(\pos)$. The Pauli matrices $\tau_{x,y,z}$ and $\sigma_{x,y,z}$ act in particle-hole space and in spin space, respectively, and we have set $\hbar=1$.

We set $\magn_\text{sk}(\pos)$ as in Eq.~\eqref{Eq:SkyrmionTexture}. The superconducting order parameter is spatially dependent to account for the vortex, $\Delta(\pos)=e^{ib\varphi}\Delta(r)$ with integer $b$ and $\Delta(r)=\Delta(1-e^{-r/R_v})$, where the vortex and the skyrmion cores are both located at the origin. This configuration can be energetically stable \cite{HSR16,BCB19,DVE18}. Then, the modified angular momentum operator
\begin{equation}\label{Eq:AngMom}
L = -i\partial_\varphi + \frac n2\sigma_z -\frac b2\tau_z
\end{equation}
commutes with the Hamiltonian, such that the eigenstates of the system can be separated into angular and radial parts, $\Psi^l(r,\varphi)=\Psi^l_\varphi(\varphi)\Psi^l_r(r)$, where $l$ denotes the eigenvalue of $L$. Following from Eq.~\eqref{Eq:AngMom}, the angular part has the form
\begin{equation}
\Psi^l_\varphi(\varphi) = e^{i\varphi\left(l-\frac n2\sigma_z+\frac b2\tau_z\right)}\,,
\end{equation}
such that $\Psi^l_\varphi(\varphi)=\Psi^l_\varphi(\varphi+2\pi)$ imposes the constraint on $l$ that
\begin{equation}
l\in\begin{cases}\mathbb{Z}&\text{if }n+b\text{ even}\\
\mathbb{Z}+\frac12&\text{if }n+b\text{ odd.}\end{cases}
\end{equation}
Furthermore, the particle-hole operator $C=\sigma_y\tau_yK$ (with complex conjugation $K$) and $L$ have the commutation relation $[L,C]=2LC$, and consequently $LC\Psi^l=-lC\Psi^l$. Therefore, if $\Psi^l$ is an eigenstate of $C$ (a Majorana mode), then necessarily $l=0$. This restricts the emergence of MBS to skyrmion-vortex pairs where $n+b$ is even, as in the realistic case $n=b=1$. Thus, in such a pair, the existence of DWS is no longer required for the emergence of MBS.

The eigenproblem for $\mathcal{H}$ can be reduced to one dimension for the radial component, $H^l(r)\Psi^l_r(r)=\epsilon^l\Psi^l_r(r)$, with
\begin{eqnarray}
\mathcal{H}^l(r) &=& -\frac{1}{2m}\left[\partial_r^2 + \frac{1}{r}\partial_r+\frac{1}{r^2}\left(l-\frac n2\sigma_z +\frac b2\tau_z\right)^2\right]\tau_z \notag\\
&&-\mu\tau_z + \lambda\sigma_z\cos g(r) + \lambda\sigma_x\sin g(r) + \Delta(r)\tau_x\,. \notag \\ 
\label{Eq:RadHamiltonian}
\end{eqnarray}
Because the radial probability density is $r|\Psi^l_r(r)|^2$, it is convenient to consider $\Phi^l(r)=\sqrt{r}\Psi^l_r(r)$. The Hamiltonian acting on $\Phi^l(r)$ reads
\begin{equation}\label{Eq:RadProbHamiltonian}
\tilde{\mathcal{H}}^l(r) = \frac{1}{\sqrt{r}}\left(\mathcal{H}^l(r) + \frac{1}{2mr}\partial_r\tau_z - \frac{1}{8mr^2}\tau_z\right).
\end{equation}

The minimal Hamiltonian~\eqref{Eq:Hamiltonian} could contain further terms that we have not taken into account so far. The presence of skyrmions implicitly requires SOC, which is missing in the Hamiltonian. Importantly, in the presence of SOC, the reasoning above remains valid only if the commutation with $L$ is preserved. This can be achieved if the usual Rashba Hamiltonian
\begin{eqnarray}
\mathcal{H}_R &=& -2i\alpha\tau_z\left(\sigma_x\partial_y - \sigma_y\partial_x\right)\notag\\
&=& -2i\alpha\tau_z e^{-i\varphi\sigma_z}\left(-\sigma_y\partial_r + \frac{1}{r}\sigma_x\partial_\varphi\right)
\end{eqnarray}
is replaced by a generalized SOC
\begin{equation}\label{Eq:Hsoc}
\mathcal{H}_\text{soc} = -2i\alpha\tau_z e^{-in^\prime\varphi\sigma_z}\left(-\sigma_y\partial_r + \frac{1}{r}\sigma_x\partial_\varphi\right)
\end{equation}
in which $n^\prime$ is chosen equal to $n$. Equation~\eqref{Eq:Hsoc} describes a SOC in which the spin direction winds $n^\prime$ times as one encircles the origin. In result, it varies with the polar angle between the Rashba and Dresselhaus types whenever $n^\prime\neq 1$. In particular, $\mathcal{H}_\text{soc}$ is identical to the SOC considered in Ref.~\onlinecite{YSK16} for $n^\prime=2$.~~\footnote{Up to a momentum-independent term that also commutes with $L$. The SOC of Ref.~\onlinecite{YSK16} can be written in polar form as $e^{-i\varphi\sigma_z}\mathcal{H}_R + \mathcal{H}_Re^{i\varphi\sigma_z}$.} Unconventional SOC with a vortex-like structure was also suggested to lead to MBS in cold-atom systems in Ref.~\onlinecite{STF09}. However, in that case only a complex phase winding $e^{im\varphi}$ with integer $m$ was introduced in the prefactor of the Rashba Hamiltonian, instead of a spin rotation, which is different from $\mathcal{H}_\text{soc}$. For arbitrary $n^\prime$, Eq.~\eqref{Eq:Hsoc} represents a rather artificial type of SOC which is compatible with a skyrmion of the same winding number. Contrarily, in our proposal of a skyrmion-vortex pair with $n=b=1$, the SWS demands that also $n^\prime=1$, such that $\mathcal{H}_\text{soc} = \mathcal{H}_R$ recovers the ordinary Rashba Hamiltonian. After the reduction to the one-dimensional radial probability $\Phi^l(r)$, $\mathcal{H}_\text{soc}$ takes the form
\begin{equation}\label{Eq:RadSOCHamiltonian}
\tilde{\mathcal{H}}_\text{soc}^l(r) = \frac{\alpha}{\sqrt{r}}\left[2i\sigma_y\tau_z\partial_r + \frac{n^\prime-1}{r}i\sigma_y\tau_z + \frac{2l}{r}\sigma_x\tau_z + \frac{b}{r}\sigma_x\right]
\end{equation}
and can be added to Eq.~\eqref{Eq:RadProbHamiltonian}. In total, electrons will be subject to the sum of the background SOC of Eq.~\eqref{Eq:Hsoc} and the synthetic SOC generated by the spatially varying magnetization. We note that the criterion for the existence of MBS is independent of the presence or absence of (commutation-preserving) background SOC. The significance of $\mathcal{H}_\text{soc}$ is to reduce the overlap of the two Majorana modes at the skyrmion core and the rim of the system. To obtain the same effect based on synthetic SOC at $\alpha=0$, the skyrmion must have multiple spin flips in the radial direction \cite{YSK16}. Numerical results on the MBS will be shown in the next section.

Another remark is at hand concerning the role of SOC. One canonical example for the formation of localized MBS is a vortex in a two-dimensional spinless $p$-wave superconductor \cite{Ali12, Ben13, Iva01}, which may effectively result from $s$-wave pairing, Rashba SOC, and Zeeman splitting. In our language, such a vortex would correspond to $n=0$ and $b=1$, where obviously $n+b$ is odd. Yet, our criterion does not contradict the well-known results about MBS in that case. The reason is the requirement of Rashba SOC, which breaks the commutation with $L$ for $n=0$, such that our criterion is not valid. For a uniform out-of-plane Zeeman field~\footnote{Note that this is a stronger requirement than $n=0$, for which a non-winding in-plane component would still be allowed.} without skyrmions, one can formulate a different criterion, namely, for a SOC of the general form Eq.~\eqref{Eq:Hsoc} and a vorticity $b$, the operator $L^\prime=-i\partial_\varphi + \frac{1}{2}n^\prime\sigma_z - \frac{1}{2}b\tau_z$ commutes with the Hamiltonian. Though formally similar to $L$, $L^\prime$ is constructed based on the winding number of the SOC instead of a skyrmion. In analogy to the arguments presented above, the Majorana criterion $(n^\prime+b)\,\text{mod}\,2\overset{!}{=}0$ can then be derived. For Rashba SOC, where $n^\prime=1$, it follows that vortices with odd $b$ will harbor MBS (similar to the result in Ref. \onlinecite{LWH18}). The canonical case is recovered with $b=1$. Whenever a skyrmion and SOC coexist, they are only compatible in terms of commutation relations if $n=n^\prime$ (then, indeed $L=L^\prime$). Otherwise, the system does not have any rotational symmetry that could give rise to a localized MBS. We will present an example of such a case at the end of the next section.

As an interesting side remark, in Majorana systems based on either $L$ or $L^\prime$, equivalent formulations can often be found by suitable transformations. One can check that, once the Hamiltonian is reduced to the Majorana sector \cite{SLT10, YSK16}, $n=b=0$ combined with unconventional SOC where $n^\prime=0$ is equivalent to $n=2$ and $b=0$ without SOC. The latter case was solved in Ref.~\onlinecite{YSK16}, see, e.g., Fig.~5 therein for the analytical Majorana wavefunctions. In Ref.~\onlinecite{STF09}, it was shown that a vortex in the superconducting order parameter can be mapped to a complex-phase vortex in the SOC. Thus, for a uniform out-of-plane Zeeman field, the case $n^\prime=b=1$ is equivalent to $b=1$ and Rashba SOC with a vortex.

Besides SOC, we have not incorporated the electromagnetic vector potential in Eq.~\eqref{Eq:Hamiltonian}, thus orbital effects and screening supercurrents in response to the magnetic texture are not included. This is a valid approximation for the following reasons. First and foremost, any vector potential of the form $\mathbf{A}=A(r)\unitvec_\varphi$ (corresponding to the out-of-plane magnetic flux in a vortex) does not change the commutation relation $[H,L]=0$ and hence does not affect the Majorana criterion. Moreover, here we focus on a system where both the chiral magnetic layer and the superconducting film are thin. The supercurrent that emerges as a consequence of a domain wall in a superconductor-ferromagnet bilayer (we may view skyrmions as a limiting case of general domain wall structures in this regard by continuously shrinking a circular domain to a skyrmion core) will be bounded by $J_\text{max}=cM(d_F/\delta)$ \cite{BuC05}, where $c$ is the speed of light, $M$ the saturation magnetization, $d_F$ the thickness of the magnetic layer, and $\delta$ the width of the domain wall (here: related to the skyrmion radius). As we focus on $d_F\ll\delta$, where $d_F$ may correspond to monoatomic layer thickness~\cite{HDL18} whereas $\delta$ will be on the order of 10 -- 100 nanometers, we can expect that $J_\text{max}$ is insignificant in the thin-film limit. In this limit, vortices may not form spontaneously, but can still be induced by an external out-of-plane magnetic field \cite{BCB19}. When skyrmions and vortices coexist, vortex pinning to the skyrmion center is energetically favorable \cite{HSR16,BCB19,DVE18}.

\section{Results on Majorana bound states in skyrmion-vortex pairs}\label{Sec:Results}

\begin{figure*}[p]
\includegraphics[width=\textwidth]{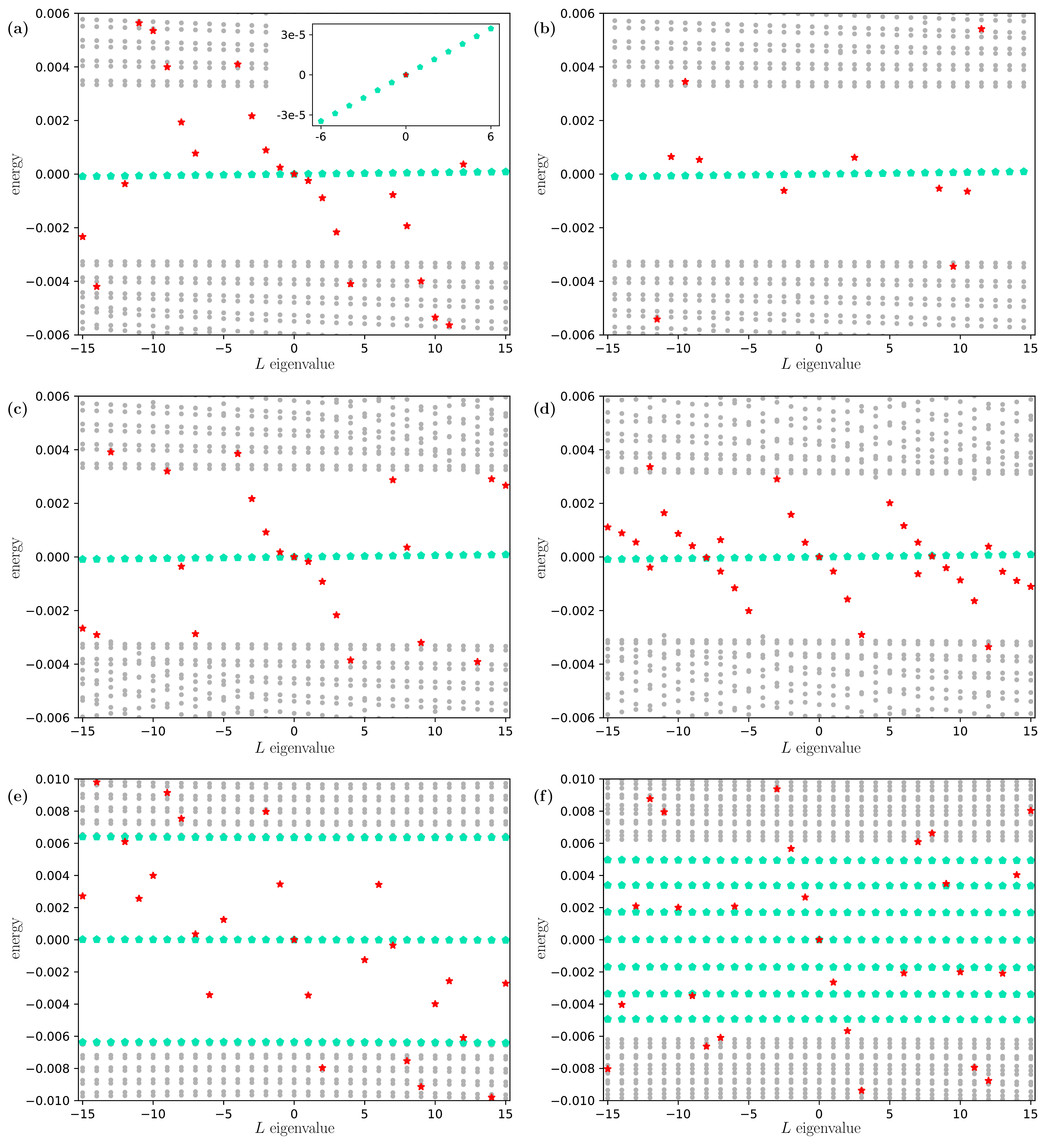}
\caption{\label{Fig:Energyplots} (a) Energy levels in a single skyrmion-vortex pair with $n=1$ and $b=1$ with the numerical parameters $\lambda=1$, $\Delta=0.5$, $\alpha=0.4$, $R=300$, $R_v=5$, $m=0.01$ solved on $N=1000$ sites in radial direction with spacing $a=1.5$. (Inset) Close-up view at very low energies revealing a linear dispersion of the rim modes. [(b)--(f)] Energy levels in a system similar to (a) with the following differences: (b) bare skyrmion without vortex ($b=0$); (c) extended vortex, $R_v=200$; (d) $R_v=400$ and $R=100$; (e) $\alpha=0$, but effective SOC from $25$-fold radial winding of $\magn$, and the system terminates at the rim of the skyrmion; (f) same as (e), but including an annulus of uniform magnetization around the skyrmion. In all panels, red stars indicate states localized near the core, whereas turqouise pentagons indicate states localized near the rim of the system (in (f) in the surrounding region outside the skyrmion).}
\end{figure*}

\begin{figure*}[t]
\includegraphics[width=\textwidth]{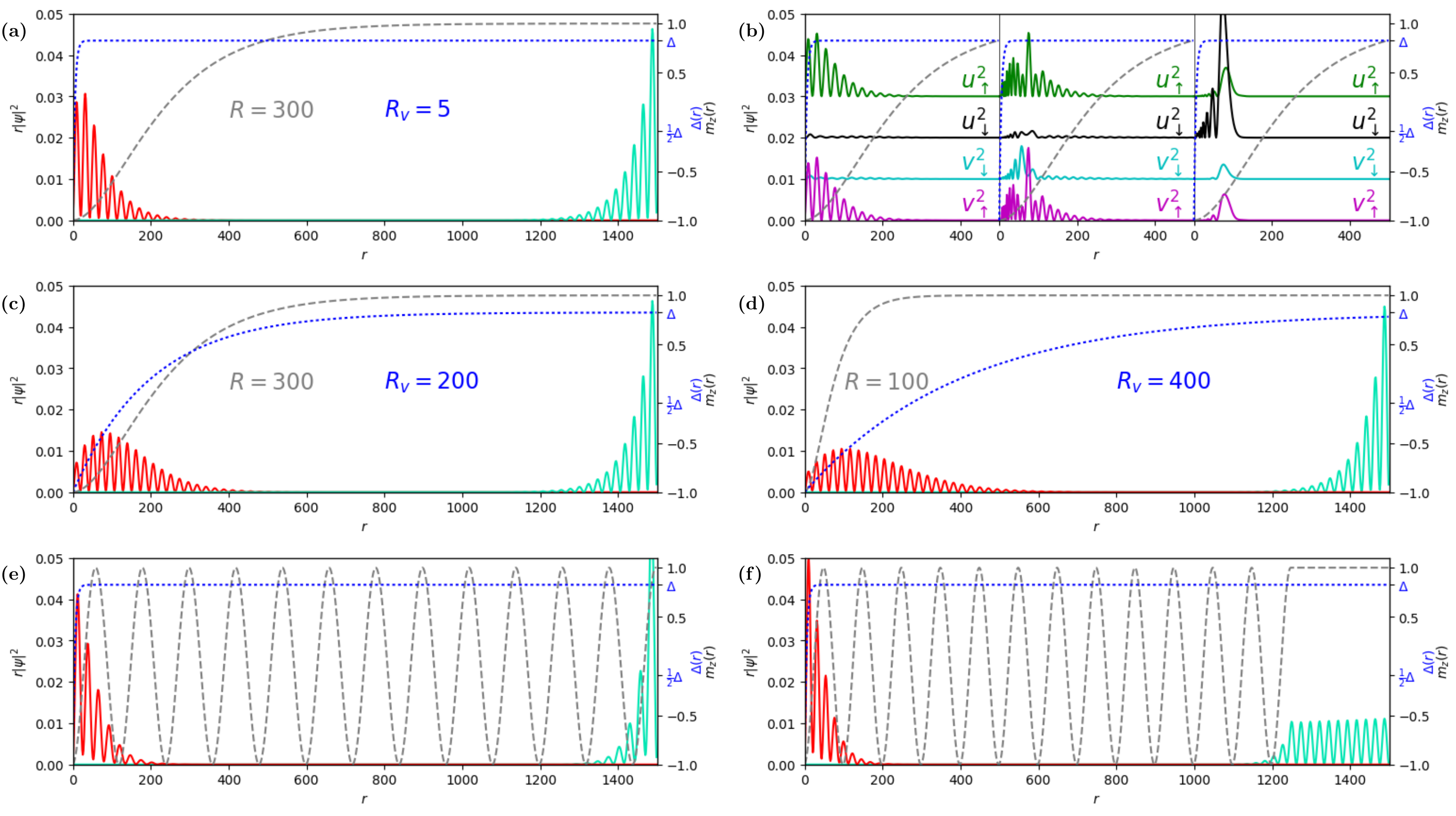}
\caption{\label{Fig:Stateplots} Radial probability density (left $y$ scale) of the inner (red solid line) and outer (turqouise solid line) Majorana modes as well as the radial shape of the skyrmion texture in terms of $m_z$ (dashed grey line, black right $y$ scale) and the profile of the vortex (blue dotted line, blue right $y$ scale). Panels: (a) MBS in the case of Fig.~\ref{Fig:Energyplots}(a); (b) the four components of the inner MBS from panel (a) (left) in comparison to two examples of localized core states at non-zero energies (i.e., not MBS) in the same skyrmion-vortex pair (middle: state at $l=-10$, right: state at $l=-12$), with different offsets added to the components for distinguishability; [(c)--(f)] MBS in the cases shown in Figs.~\ref{Fig:Energyplots}(c)--\ref{Fig:Energyplots}(f), respectively. With an extended vortex, (c) and (d), the localization of the inner MBS becomes weaker. In (f), the outer mode is delocalized between the outer skyrmion radius and the system rim.}
\end{figure*}

Now we present our results obtained for the eigenstates in skyrmion-vortex pairs by exact diagonalization of Eq.~\eqref{Eq:RadProbHamiltonian} [together with Eq.~\eqref{Eq:RadSOCHamiltonian} whenever $\alpha\neq0$]. The low-energy spectra are shown in Fig.~\ref{Fig:Energyplots}. Our main result concerns the case with realistic winding numbers $n=1$ and $b=1$ and an exponential radial shape of both the skyrmion and the vortex in the presence of background SOC [Fig.~\ref{Fig:Energyplots}(a)]. The skyrmion is chosen to be of N\'eel type. A pair of zero-energy states is found at $l=0$, where one of these states is localized at the core of the skyrmion, whereas the other one is located at the rim of the system. These are the MBS predicted by the criterion in the previous section. Thus, it is possible to nucleate a MBS in a skyrmion without a demand for higher winding numbers. The probability densities of the two zero modes are shown in Fig.~\ref{Fig:Stateplots}(a).

Apart from the MBS, this system has further noteworthy features. First, a band of close-to-zero energy states appears at any eigenvalue of $L$. All of them are located at the rim of the system. A close-up of the spectrum (inset in Fig.~\ref{Fig:Energyplots}(a)) reveals that this band exhibits a weak linear dispersion, thus none of the states at $l\neq0$ are truly at zero energy. Rather, this band represents a chiral Majorana mode at the rim, where the tiny slope of the band with respect to angular momentum yields the group velocity when multiplied with the system radius. The presence of such a chiral mode is consistent with the results of Ref. \onlinecite{ChS15}. A second feature is the emergence of further bound states at the core besides the MBS. In Fig.~\ref{Fig:Stateplots}(b), two examples of such states are compared to the MBS. The components $(u_\uparrow, u_\downarrow, v_\downarrow, -v_\uparrow)$ of a MBS wavefunction generally have to satisfy the conditions $u_\uparrow=c^*v^*_\uparrow$, $u_\downarrow=c^*v^*_\downarrow$, where $c$ with $|c|=1$ is the eigenvalue of the particle-hole operator $C$. In contrast, all components are independent for the non-Majorana bound states. These additional bound states exist for two reasons. First, it is expected \cite{PNB16, PWP16} that a skyrmion on a superconductor will generally induce bound states which relate to Yu-Shiba-Rusinov states in the limit of a very small skyrmion (a tiny skyrmion is conceptionally similar to a magnetic impurity). Second, it is known that also a vortex line in a superconductor causes bound states \cite{CGM64}. Interestingly, the core states close to $l=0$ are reminiscent of a band, signaling an emergent chiral character. The slope has opposite sign compared with the band of rim states. The core and rim bands can be interpreted as the inner and outer chiral modes of a topological superconductor with the geometry of an annulus, where the hole stems from the vortex. In our calculations, bound states exist at various energies both above and below the effective superconducting gap. The localization of states outside the gap is not necessarily exponential. In Fig.~\ref{Fig:Energyplots}, we have used the qualitative criterion that $70\,\%$ of the total probability of a state must be accumulated on $20\,\%$ of the radial extent of the system for this state to be classified as localized.

For comparison, Fig.~\ref{Fig:Energyplots}(b) shows the spectrum obtained from a bare skyrmion in the absence of a vortex, i.e., $b=0$. No MBS are present. Note that $l$ takes half-integer values in this case, which is similar to the reversed situation $n=0$ and $b=1$. The outer chiral mode persists, but without a state at exactly zero energy. Also, localized states at the core are still present, though not as many as in the presence of a vortex, and without the formation of a band-like structure.

Next, we investigate the impact of the vortex radius, which will be of practical importance for experiments. To obtain MBS from skyrmion-vortex pairs, it is generally desirable that the vortex is small compared to the skyrmion: If electrons inside the skyrmion would essentially not feel a pairing potential because the vortex is too big, it is clear that the joint effect of the skyrmion texture and superconductivity would not be observable. In Fig.~\ref{Fig:Energyplots}(a), $R_v$ was therefore set to $5$ compared to the skyrmion radius $R=300$. In panels (c) and (d), the radii are $R_v=200$, $R=300$ and $R_v=400$, $R=100$, respectively, with otherwise identical parameters. In both cases, MBS are still found, thus they persist over a large range of relative sizes of the skyrmion compared to the vortex. However, the localization of the inner MBS becomes weaker because of the reduced superconducting gap close to the core, see Figs.~\ref{Fig:Stateplots}(c) and \ref{Fig:Stateplots}(d). In consequence, the overlap between the inner and outer MBS increases, thus the numerically obtained energy in the last case, $\epsilon_\text{MBS}\approx 8\times10^{-8}$, is an order of magnitude larger than in the first panel (small $R_v$), where $\epsilon_\text{MBS}\approx 6\times10^{-9}$.

For comparison with the earlier results of Ref.~\onlinecite{YSK16}, we have also considered skyrmion-vortex pairs in which the effective SOC stems from a winding of the magnetization along the radial direction, while $\alpha=0$. More specifically, $g(r)=\pi(1-r/R)$ if $r<25 R$ and $g(r)=0$ otherwise, with $R=30$ (thus, the outer radius of the skyrmion texture is $25R$). In Fig.~\ref{Fig:Energyplots}(e), the system is restricted to the skyrmion, whereas the skyrmion is embedded in a region of uniform magnetization in panel (f). In agreement with Ref. \onlinecite{YSK16}, the radial winding of the magnetization is suitable to stabilize MBS. Panel (e) is reminiscent of the results in Ref. \onlinecite{YSK16}, with a MBS at the core and one at the rim of the skyrmion. Notably, if the skyrmion has a surrounding, the outer mode is not localized at the rim of the system, but rather delocalized between the outer radius of the skyrmion and the rim of the system, cf. Figs.~\ref{Fig:Stateplots}(e) and \ref{Fig:Stateplots}(f). This is ascribed to the absence of background SOC. Here, the effective SOC is only present inside the skyrmion, and no SOC emerges otherwise. In contrast, in the previously discussed cases with $\alpha\neq0$, the uniform background SOC enforces the outer modes to reside at the system rim. This difference does not become apparent as long as the system size is restricted to the skyrmion itself. In addition, the surrounding of the skyrmion harbours numerous further bound states at energies below the effective gap, which form multiple weakly dispersive bands at different energies. The number of such states can be expected to increase further for larger system sizes. As in all other panels, multiple bound states near the core appear at various energies inside and outside the effective gap. We have also computed energy spectra for DWS without a vortex ($n=2$, $b=0$, not shown), which are qualitatively similar to panels (e) and (f).

\begin{figure*}
\flushleft{(a)\hspace{\columnwidth}(b)}\\
\includegraphics[width=0.95\columnwidth]{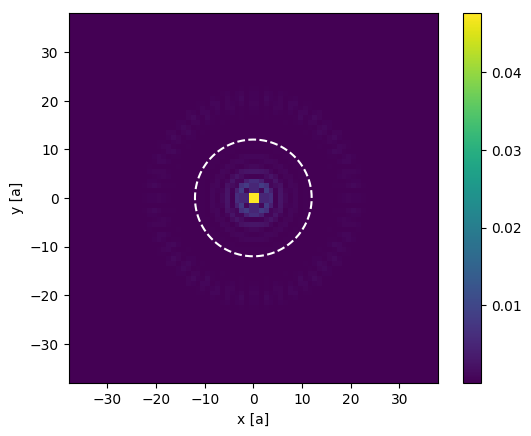}\hfill
\includegraphics[width=0.9822\columnwidth]{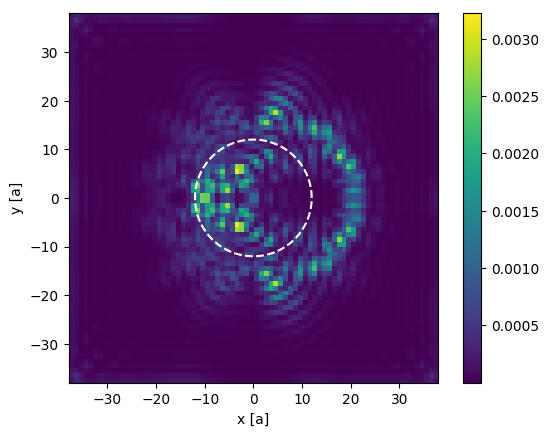}
\caption{\label{Fig:Rashba} The probability density of the lowest-energy states in two dimensions for (a) a skyrmion-vortex pair of $n=b=1$, where Rashba SOC commutes with the modified angular momentum operator $L$, and (b) for a DWS without a vortex, $n=2$, $b=0$, where Rashba SOC breaks the commutator. The white dashed line indicates the skyrmion radius $R$.}
\end{figure*}

Finally, we discuss the impact of SOC with respect to the rotational symmetry of the system, as expressed by the commutation with the operator $L$, cf. Eq.~\eqref{Eq:AngMom}. If the symmetry is broken by a mismatch of $n$ and $n^\prime$, the reduction to a one-dimensional system fails. Therefore, we now diagonalize the Hamiltonian numerically in two dimensions. We consider two cases: (i) a skyrmion-vortex pair with $n=b=1$ and Rashba SOC as before, where $[\mathcal{H},L]=0$, and (ii) a DWS in the absence of a vortex ($n=2$, $b=0$) with Rashba SOC, such that $n\neq n^\prime$ and $[\mathcal{H}_R, L]\neq0$ breaks the commutation relation. In Fig.~\ref{Fig:Rashba}, the probability density of the lowest-energy state is shown for both systems. In the first case, a symmetrically localized core state near zero energy is found, which is consistent with the existence of MBS in this case (only the inner state is shown in the figure) and qualitatively confirms our previous results. We note, though, that finite-size effects are much larger than in the one-dimensional calculations, owed to computational limitations in two dimensions. The energy of the MBS is therefore still $\approx 8\times10^{-5}$ times the hopping amplitude in our calculations. In the second case, the lowest-energy state clearly reveals the broken symmetry. It is also not well-localized anymore and raises in energy. This state and its partner state (related through particle-hole symmetry $\mathcal{C}$) do not form a pair of spatially separated inner and outer states, as expected for MBS in skyrmions. These results demonstrate that the unconventional form of the background SOC $\mathcal{H}_\text{soc}$ in Eq.~\eqref{Eq:Hsoc} is important when $n\neq 1$ and cannot be substituted by Rashba SOC. On the other hand, the compatibility of the skyrmion-vortex pair with $n=1$ with ordinary Rashba SOC represents a further advantage of our proposal compared to skyrmions of higher winding numbers. %
In Fig.~\ref{Fig:Rashba}, we used a square lattice with $76\times76$ sites and the numerical parameters, in units of the nearest-neighbor hopping, $\mu=0$, $\alpha=0.8$, $\Delta=0.5$, and a skyrmion radius of $R=12a$ (vortex radius $R_v=R/10$ in the first case).

\section{Summary}\label{Sec:Conclusion}
We have studied possibilities to obtain localized Majorana bound states (MBS) from skyrmions at a superconductor-chiral magnet interface. We have shown that the formation of skyrmions with winding number two, i.e. DWS, in in such systems is energetically unfavorable, even if one can enforce a merger of two circular skyrmions of winding number one. Instead, one of these skyrmions is deleted through a point-like discontinuity. Therefore, DWS as MBS carriers seem unrealistic.

Nevertheless, we have demonstrated that a composite object consisting of a skyrmion and a concentrical vortex can support a localized MBS at its core, if the winding numbers of the skyrmion and the vortex add up to an even number. This allows the creation of MBS with experimentally accessible skyrmions of winding number one. Importantly, this case is also compatible with ordinary Rashba spin-orbit coupling, which would break the rotational symmetry of the system otherwise. We have calculated the Majorana energies and their wavefunctions, see Fig.~\ref{Fig:Energyplots} and \ref{Fig:Stateplots}, showing the localization on the radial axis and the spatial separation of the inner and the outer mode. MBS at skyrmion cores indicate a route towards direct spatial control of non-Abelian quantum states in two dimensions.

We hope that our proposal will be realized in future experiments. Already now, magnetic bilayer systems hosting skyrmions are available. In addition, there is ongoing progress towards the observation of skyrmions below the superconducting transition temperature in such bilayers \cite{Private}. In order to generate skyrmion-vortex pairs rather than bare skyrmions, the superconducting layer should preferably be a type-II superconductor. Alternatively, this layer could be fabricated sufficiently thin compared to the penetration depth of the material.

We expect that future work will address the stability of the MBS under distortions from the circular shape of the skyrmions. Shape distortions can be caused, e.g., by the geometry of the sample \cite{HDL18, JLK17}, or by dynamic oscillatory modes which might be activated when operations are performed with the skyrmions. Furthermore, real samples will harbour multiple skyrmions or skyrmion-vortex pairs. Hence, hybridization effects of the electronic states bound to different skyrmion-vortex pairs should be studied, as well as many-skyrmion arrangements in general.

\textsl{Note added:} Recently, we learned about Ref.~\onlinecite{GMS19}, which also addresses topological superconductivity induced by magnetic skyrmions.

\begin{acknowledgments}
We thank Dmitry Aristov, Stefan Bl\"ugel, Igor Burmistrov, Marie Herv\'e, Markus Garst, Leonid Glazman, Jens Paaske, and Wulf Wulfhekel for useful and inspiring discussions. The work was supported by the Deutsche Forschungsgemeinschaft via the grants MI 658/9-1 (joint DFG-RSF project) and MI 658/7-2 (Priority Programme 1666 ``Topological Insulators''). I.V.G. acknowledges support by the Russian Science Foundation through grant No. 17-12-01182.
\end{acknowledgments}

\appendix

\section{Dzyaloshinskii-Moriya energy of skyrmions}\label{App:DMEnergy}
In this appendix, we provide a proof for the two statements from Sec.~\ref{Sec:Merger} of the main text that (i) the Dzyaloshinskii-Moriya contribution to the energy vanishes in a DWS, and (ii) the rotation of the magnetic texture in a SWS (by alteration of $\varphi_0$) is counteracting the Dzyaloshinskii-Moriya interaction, leading to an energy barrier proportional to $DR$.

For slowly varying magnetization (compared to the lattice constant $a$), $\magn(\pos)$ can be expanded at site $i$,
\begin{equation}
m_\beta(\pos) = m_\beta(\pos_i) + (\pos-\pos_i)\cdot\boldsymbol{\nabla}m_\beta(\pos_i),
\end{equation}
($\beta=x,y,z$) to get a continuum version of Eq.~\eqref{Eq:MagEnergy}. Using the vector
\begin{equation}
\mathbf{D}_{ij} = D\unitvec_z \times \frac{\left(\pos_j-\pos_i\right)}{a}
\end{equation}
acting on a pair of nearest neighbors $\left<ij\right>$, the overall Dzyaloshinskii-Moriya energy for a magnetic texture as in Eq.~\eqref{Eq:SkyrmionTexture} then takes the form
\begin{eqnarray}
E_\text{DMI} &=& 
 \frac{1}{2}\frac{D}{a}\sum_{\boldsymbol{\rho}}\int_0^\infty \!\! \frac{rdr}{a^2}\int_0^{2\pi}\! d\varphi\Big[\frac{1}{r}\left(\frac{df}{d\varphi}\right)\sin g\cos g\notag\\
 &\times&
\left(\rho_x\sin f - \rho_y\cos f\right)\left(\rho_x\sin\varphi-\rho_y\cos\varphi\right)\notag\\
&+& \left(\frac{dg}{dr}\right)\left(\rho_x\cos f + \rho_y\sin f\right)\left(\rho_x\cos\varphi + \rho_y\sin\varphi\right)\Big]. \notag \\
\end{eqnarray}
Here, we denote by $\boldsymbol{\rho}$  all possible nearest-neighbor bond vectors in a given lattice type. The factor of $\frac{1}{2}$ corrects the double-counting of bonds in the integration. When $f(\varphi)=n\varphi + \varphi_0$ is inserted, the integration over $\varphi$ cancels all terms whenever $n\neq\pm1$. In particular, a DWS has $E_\text{DMI}=0$, proving statement (i). For $n=1$,
\begin{equation}
E_\text{DMI} = \frac{\pi\eta D}{2a}\cos\varphi_0\left[-R\int_{g(0)}^{g(\infty)}\!\!\!dg\,\frac{\sin(2g)}{2g}-\int_0^\infty\!\!\!dr\,g(r)\right],
\end{equation}
where we have used $\rho_x^2+\rho_y^2=a^2$ and introduced the coordination number $\eta$ of the lattice. A similar expression was found in Ref. \onlinecite{WYW18}. The first integral always yields $\frac{1}{2}\text{Si}(2\pi)\approx 0{.}71$ for any monotonous $g(r)$ with correct boundary conditions. The second integral gives $\pi R$ for our specific choice of $g(r)$. In total,
\begin{equation}
E_\text{DMI} = -\mathcal{K}\eta D\frac{R}{a}\cos\varphi_0\,.
\end{equation}
with the numerical constant $\mathcal{K}\approx6.05$. With respect to a rotation of the skyrmion texture, i.e. changing $\varphi_0$, the Dzyaloshinskii-Moriya energy is proportional to $\cos\varphi_0$, such that $\varphi_0=0$ (N\'eel), $\varphi_0=\frac{\pi}{2}$ (Bloch) and $\varphi_0=\pi$ (anti-N\'eel) correspond to minimal, zero, and maximal energy, respectively, if $n=1$, and vice versa if $n=-1$. This proves statement (ii). With the parameters used in Sec.~\ref{Sec:Merger} on the triangular lattice ($\eta=6$), one obtains that rotating the texture of one skyrmion by $\pm\frac{\pi}{2}$ changes the Dzyaloshinskii-Moriya energy in the system approximately by $1600\,\text{meV}$.

\section{Details on the merger implementation}\label{App:Merger}
Here, we describe how the continuous and discontinuous mergers of two SWS, cf. Fig.~\ref{Fig:Merger}, have been implemented to obtain the result shown in Fig.~\ref{Fig:Merger-Energy}. First, we introduce the notion of a skyrmion field $\magn_\mathrm{sk}(x,y)$, which is obtained from Eq.~\eqref{Eq:SkyrmionTexture} by substracting the uniform background $\magn_\mathrm{bg}=\unitvec_z$. %
The boundary conditions for $g(r)$ imply that the skyrmion field vanishes at infinity. We denote the SWS and DWS fields $\magn_\text{SWS}$ and $\magn_\text{DWS}$, respectively. A configuration with more than one skyrmion is now obtained by summing $\magn_\textrm{bg}$ and multiple skyrmion fields and subsequently normalizing to $\left|\magn(\pos)\right|=1$. For the merger process, we initially place two identical N\'eel SWS (with optimal radius, cf. Fig.~\ref{Fig:EnergyRadius}) on the $x$ axis at a distance $d$, which is gradually decreased.

For the discontinuous merger, the overall magnetic texture is
\begin{equation}\label{Eq:DiscontinuousMerger}
\magn(\pos, d) = \frac{\overline{\magn}(\pos, d)}{\left|\overline{\magn}(\pos, d)\right|} \,,
\end{equation}
with the not normalized vector field
\begin{eqnarray}
&&\overline{\magn}(\pos, d) = \notag\\
&&~~(1-w)\left[\magn_\text{SWS}(x+d/2, y) + \magn_\text{SWS}(x-d/2, y)\right] \notag\\
&&{~~}+ w\,\magn_\text{SWS}(x,y) + \magn_\text{bg}\,,
\end{eqnarray}
and the weight
\begin{equation}\label{Eq:MergerWeight1}
w = \Theta(d_1-d)\left(\frac{d_1-d}{d_1}\right)^2 \,.
\end{equation}
By means of $w$, we smoothly interpolate between the superposition of two skyrmions and a single skyrmion for small $d$ ($d_1$ is specified below) in order to ensure that the final skyrmion is identical to each of the initial ones. Otherwise, the final state would also be a SWS, but with a different radial profile.

This process starts with two SWS, thus $n_\text{tot} = 2$, but terminates with a single SWS, $n_\text{tot}=1$. The change of the total winding number happens at a critical distance $d_0$, where Eq.~\eqref{Eq:DiscontinuousMerger} has a discontinuity at $(x,y)=(0,0)$. Namely, $\overline{\magn}(0, d_0)=0$, such that the normalization fails. We find that $d_0=2R\text{ln}3$ for the exponential radial profile. It is easy to show that $\overline{\magn}\neq0$ in any other case. In Eq.~\eqref{Eq:MergerWeight1}, $d_1$ was set to $0{.}8d_c$.

\begin{figure}[t]
\includegraphics[width=0.7\columnwidth]{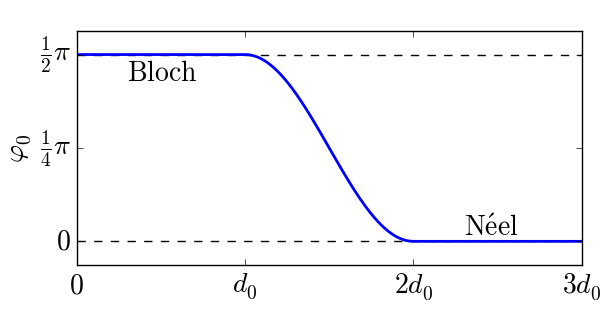}
\caption{\label{Fig:OffsetAngle}The offset angle for the left SWS as a function of the center-to-center skyrmion distance in the continuous merger. The right SWS has the same offset with negative sign.}
\end{figure}

In the continuous merger, where $n_\text{tot}$ is conserved, a point where $\overline{\magn}=0$ has to be avoided by rotating the two SWS at $d>d_0$. More precisely, the offset angle $\varphi_0$ in $f(\varphi)$ is smoothly changed from $0$ to $\pm\frac12\pi$, with $+$ ($-$) for the left (right) SWS:
\begin{equation}
f_\pm(\varphi) = \varphi \pm
\begin{cases}
0 & \text{if } d>2d_0 \\
\displaystyle\frac{\pi}{4}\left[1+\cos(\pi\frac{d-d_0}{d_0})\right] & \text{if } d_0<d\leq 2d_0 \\
\frac{\pi}{2} & \text{if } d\leq d_0
\end{cases}\,.
\end{equation}
The angles at $d<d_0$ are chosen to be precisely $\pm\frac12\pi$ in order for the resulting skyrmion to have the rotational symmetry described by $L$, Eq.~\eqref{Eq:AngMom}. We show the offset angle of $f_+(\varphi)$ in Fig.~\ref{Fig:OffsetAngle} for clarity. The respective fields are labeled $\magn_\text{SWS}^\pm$.

The vector field prior to normalization now reads
\begin{eqnarray}
&&\overline{\magn}(\pos, d) = \notag\\
&&~~(1-w)\left[\magn_\text{SWS}^+(x+d/2, y) + \magn_\text{SWS}^-(x-d/2, y)\right] \notag\\
&&{~~}+ w\,\magn_\text{DWS}(x,y) + \magn_\text{bg}\,,
\end{eqnarray}
and here we use the weight
\begin{equation}\label{Eq:MergerWeigfht2}
w = \frac12\Theta(d_1-d)\left[1-\cos\left(\pi\frac{d_1-d}{d_1}\right)\right]
\end{equation}
to smoothly transition into the final state, with $d_1=0.8d_0$ as before. The final-state skyrmion field $\magn_\text{DWS}$ is slightly shrinked in comparison to the initial SWS, $R_\text{DWS}\overset{\sim}{=}\frac23R_\text{SWS}$ to reduce the energy of the state.

The details of this specific implementation of the skyrmion mergers are to some degree arbitrary, and some optimization with respect to the energy throughout the process could be done. This effort seems futile, though, as the modification of $\varphi_0$ in the continuous case must persist. This feature is mainly responsible for the energy difference between the two scenarios, as discussed in the main text, such that most modifications of the implementation would not change the physical result, namely, the deletion of one SWS in an enforced skyrmion merger. In order to reverse the energy proportions, the skyrmion would have to be shrinked in size: the barrier in the discontinuous merger is approximately constant, whereas it is proportional to $R$ in the continuous merger (see previous appendix). Figure~\ref{Fig:Merger-Energy} would suggest to consider skyrmions which are roughly five times smaller in order to equalize the energy cost in both processes. In practice, however, $R$ is not an independent parameter but has a nontrivial dependence on the coefficients in Eq.~\eqref{Eq:MagEnergy}. Therefore, such an estimate cannot be extracted from our results.

\bibliography{Majorana-Skyrmion-v6}

\end{document}